\documentclass[conference]{IEEEtran}

\usepackage[english]{babel}

\usepackage{amsmath}
\usepackage{amssymb}
\usepackage{amsfonts}
\usepackage{graphicx}
\usepackage{float}
\usepackage{parskip}
\usepackage{comment}
\usepackage{csquotes}
\usepackage[colorlinks=true, allcolors=blue]{hyperref}
\usepackage[
backend=biber,
style=ieee,
sortcites=true,
]{biblatex}
\usepackage[compact]{titlesec}

\newcommand{\flopj}{\, \textrm{FLOP}/\textrm{J}}

\newcommand{\watt}{\, \textrm{W}}

\newcommand{\nm}{\, \textrm{nm}}
\newcommand{\mm}{\, \textrm{mm}}

\newcommand\blfootnote[1]{%
  \begingroup
  \renewcommand\thefootnote{}\footnote{#1}%
  \addtocounter{footnote}{-1}%
  \endgroup
}

\usepackage{circuitikz}

\title{Limits to the Energy Efficiency of CMOS Microprocessors}
\author{Anson Ho$^*$, Ege Erdil$^*$, Tamay Besiroglu$^{*\dagger}$}

\begin{document}

\maketitle
\thispagestyle{plain}
\pagestyle{plain}

\begin{abstract}
CMOS microprocessors have achieved massive energy efficiency gains but may reach limits soon. This paper presents an approach to estimating the limits on the maximum floating point operations per Joule (FLOP/J) for CMOS microprocessors. We analyze the three primary sources of energy dissipation: transistor switching, interconnect capacitances and leakage power. Using first-principles calculations of minimum energy costs based on Landauer's principle, prior estimates of relevant parameters, and empirical data on hardware, we derive the energy cost per FLOP for each component. Combining these yields a geometric mean estimate of $4.7 \times 10^{15} \, \, \textrm{FP4/J}$ for the maximum CMOS energy efficiency, roughly two hundred-fold more efficient than current microprocessors.
\end{abstract}

\blfootnote{This paper has been accepted for publication in the 2023 IEEE International Conference on Rebooting Computing. \copyright 2023 IEEE. Personal use of this material is permitted. Permission from IEEE must be obtained for all other uses, in any current or future media, including reprinting/republishing this material for advertising or promotional purposes, creating new collective works, for resale or redistribution to servers or lists, or reuse of any copyrighted component of this work in other works.}
\blfootnote{$^*$Epoch, $^\dagger$MIT FutureTech. Our code is made available in \href{https://colab.research.google.com/drive/18WKCe6QrZ8qCA5v0pZGaNhq7ovCXmqFL?usp=sharing}{this colab notebook}, and relevant data is available in \href{https://docs.google.com/spreadsheets/d/1hM-XCJ2Yob-Ne1hHv9164Px8spMxn-HcT5Fj8qSrx5E/edit?usp=sharing}{here}.}

\section{Introduction}
Driven by Moore's law and Dennard scaling, digital Complementary Metal-Oxide Semiconductor (CMOS) devices have seen massive improvements in energy efficiency over the past few decades. This is perhaps best illustrated by Koomey's Law \cite{Koomey2011ImplicationsOH}, which states that the Floating Point Operations (FLOP) per Joule dissipated doubled once every 1.5 years between 1946 and 2000 \cite{Koomey2011ImplicationsOH}, and every 2.7 years post-2000 \cite{Koomey2015energyEfficiencyUpdate}.
More recently, \cite{Hobbhahn2023TrendsMLhardware} finds that GPUs with float32 number formats have had an energy efficiency doubling time of about 2.7 years over the last 15 years.
But how far can these energy efficiency improvements continue before technology scaling hits physical limits?

Despite its practical importance, research into this particular question has been limited thus far. 
Some near-term forecasts of energy efficiency exist—for example, the ``More Moore" chapter of the 2022 IEEE International Roadmap for Devices and Systems (IRDS) report forecasts the operations per second per Joule until 2037 \cite{IEEE2022IRDSmoreMoore}. 
However, the report focuses on projections over the next decade rather than on the limits to the FLOP/J achievable by CMOS processors.
There have been relatively fewer attempts to derive these fundamental limits more directly, such as \cite{Agarwal2016EfficiencyLimitsLogicMemory} and \cite{Frank2005ApproachingTP}. 
However, these papers primarily focus on the energy costs from logic (i.e. switching transistors) and tend to neglect the costs from switching interconnect capacitances, which can be the main source of dynamic energy dissipation in modern CMOS devices (see section \ref{sec:energy-cost-accounting}). 

Some previous estimates of efficiency limits, such as those in IRDS reports, lack clearly explained methodologies and uncertainty ranges. This makes the forecasts difficult to rigorously critique and improve upon. For example, \cite{Frank2005ApproachingTP} predicted that ``current [2005] technology" would have a maximum performance per watt of around 10 GigaFLOP per second at FP64 precision. However, without quantified uncertainties, it is unclear if later efficiency improvements in GPUs agree with this forecast. The lack of formal uncertainty ranges prevents rigorous evaluation of the accuracy of such predictions. Our work addresses this by using transparent calculations and providing uncertainty estimates. This enables constructive critiques to iteratively enhance our model.

The goal of this work is to estimate an upper bound to the energy efficiency in FLOP/J of CMOS microprocessors that is accurate to an order of magnitude. We approach this by applying standard techniques and results from microelectronics and technological forecasting and make two main novel contributions. First, we estimate upper bounds to the energy efficiency of CMOS processors \emph{based on interconnect dissipation}. Second, we extend the analysis from \cite{Miller2016AttojouleOF, Frank2005ApproachingTP, Agarwal2016EfficiencyLimitsLogicMemory} and perform an accounting of the key energy costs in the limit of CMOS microprocessors that are maximally optimized for energy efficiency. Throughout, we provide uncertainty ranges for each of our estimates and try to appropriately account for the uncertainty over each of the relevant parameters.

\section{Background}

\subsection{Landauer limit}
One theoretical FLOP/J upper bound that applies to CMOS processors comes from Landauer's principle—this posits that irreversible operations fundamentally release energy $E \geq k_B T \ln 2$, where $T$ is temperature and $k_B$ is the Boltzmann constant \cite{Landauer1961IrreversibilityAH}. Given data on the number of bit erasures per floating point operation (FLOP), Landauer's limit allows estimation of the maximum achievable FLOP per Joule (FLOP/J). We refer to such a bound, which assumes that energy dissipation via irreversible operations are the relevant energy efficiency bottleneck, as the ``Landauer limit".

\subsection{Energy cost accounting}
\label{sec:energy-cost-accounting}

\begin{table*}[h]
    \renewcommand{\arraystretch}{1.15} 
    \centering
    \begin{tabular}{c|l|c|c}
      \textbf{Category} & \textbf{Energy source} & \textbf{Energy cost} & \textbf{References}\\
      \hline
      5 (Storage) & Storing a bit in DRAM & 10 fJ & \cite{Miller2016AttojouleOF} \\
      4 (Static) & Leakage current per transistor (100 ps)\footnotemark[2] & $\sim$1 aJ & \cite{Rjoub2020AccurateLC} \\
      3 (Short circuit) & Short circuit (inverter) & 20 fJ & \cite{Wiltgen2013PowerStaticCMOS} \\
      1 (Logic) & Switching a CMOS gate & 100 aJ—100 fJ & \cite{Miller2016AttojouleOF, Manipatruni2018ScalableEM, Nikonov2015BenchmarkingOB}\\
      1 (Logic) & Floating Point Operation (fp16) & $\sim$150 fJ & \cite{Anderson2023OpticalT, Hobbhahn2023TrendsMLhardware}\\
      2 (Interconnect) & Communicating a bit across a chip & 600 fJ & \cite{Miller2016AttojouleOF}\\
      2 (Interconnect) & DRAM memory access (per bit) & $\sim$5 pJ & \cite{Horowitz2014ComputingsEnergyProblem, Keckler2011GPUsAT} \\
      2 (Interconnect) & Communicating a bit off-chip & 1--10 pJ & \cite{Miller2016AttojouleOF} \\
    \end{tabular}
    \caption{Energy costs for different operations in modern microprocessors.\protect\footnotemark[3]}
    \label{tab:energy-costs}
\end{table*}

{ 
  \footnotetext[2]{This cost is calculated over a time period of 100 ps, which is approximately the time for a transistor to switch \cite{Stellari2006SwitchingTE}.}
  \footnotetext[3]{Note that specific numbers are likely to vary depending on the hardware and task; the presented numbers are merely meant to provide an illustration of the most important energy costs at present.}   
  \setcounter{footnote}{3}
}

Energy dissipation in CMOS devices comes from two contributors: dynamic dissipation and static dissipation \cite{Ng22powerCMOS}. 
The former is due to ``dynamic" switching operations to change logic states or communicate information. 
The latter occurs in the absence of switching (``static"), and is instead primarily due to leakage currents  \cite{Ng22powerCMOS, Wiltgen2013PowerStaticCMOS}, such as charge carrier tunneling through gate dielectrics \cite{Ranurez2006gateTunnelingCurrentMOS}. 

We decompose the key sources of energy dissipation into five main categories, which we illustrate in Table \ref{tab:energy-costs}: (1) switching transistors, (2) switching interconnect wire capacitances, (3) short circuit power dissipation, (4) static power, and (5) information storage. 

\textbf{Dynamic power}:
The first three of these are examples of dynamic power dissipation.\footnote{Note that we consider multiple elements in the table as contributing to the same broad energy source—e.g. the energy costs for DRAM memory access may involve charging off-chip interconnect wires \cite{Vogelsang2010UnderstandingDRAMenergy, Miller2016AttojouleOF, Sze2017EfficientProcessingDNNs, Horowitz2014ComputingsEnergyProblem}, so we consider this as a contributor to ``switching interconnect wire capacitances".} 
Historically, the contribution from interconnects has been growing over time.
In particular, Magen \textit{et al.} (2004) find that roughly 50\% of the dynamic power dissipation was dominated by wires \cite{Magen2004InterconnectpowerDI}, and it appears likely that interconnect contributions have come to dominate power dissipation.

The primary reason for this is that power dissipated per transistor is proportional to the square of transistor linear dimensions, and with continued scaling under Moore's Law, the power use per transistor has decreased \cite{Koomey2011ImplicationsOH}.\footnote{Note that this does not tell us whether or not the overall contribution of transistors to an integrated circuit's power dissipation has increased or decreased \textit{per se}, since the decrease in transistor size has been accompanied by an increased quantity of transistors on chips.} 
On the other hand, the total length of interconnect wires on chips has been increasing—with kilometers of interconnect on modern ICs \cite{Moiseev2015AnOverviewVLSIinterconnect}—while the capacitance per unit length of interconnects are largely independent of scale \cite{Miller2016AttojouleOF}. This results in a greater total wire capacitance that needs to be charged per FLOP, such that interconnect has increasingly dominated dynamic power dissipation.

\textbf{Eliminating energy sources from the model}:
For completeness, we also include a fifth category for information storage, but the energy costs from this are much smaller than from static or dynamic dissipation and are consequently ignored.

Besides the energy costs for storage, we can also \textit{a priori} eliminate short circuit power dissipation as a key source of energy dissipation in our model. 
This only occurs when there is a direct path current from the supply to ground, which happens during a fraction of the time for switching CMOS gates. 
The short circuit power is thus typically small and transient compared to e.g. transistor switching costs \cite{Ng22powerCMOS}, and thus are unlikely to affect our conclusions by more than a small ($\leq 1.5\times$) multiplicative factor.\footnote{For instance, \cite{Wiltgen2013PowerStaticCMOS} finds 20 fJ for an inverter, compared to 100 fJ for charging and discharging. The paper also points out that short circuit contributions to energy dissipation have been decreasing over time.}
This is swamped by our uncertainty in the energy costs from e.g. interconnect dissipation, which can range over orders of magnitude (OOMs), and thus we do not explicitly account for short circuit power in our model. 

\textbf{Static power}:
It is tempting to eliminate static power dissipation \textit{a priori} as well, especially given the very low energy cost due to leakage currents per transistor. 
For instance, \cite{Rjoub2020AccurateLC, Khanna2016ShortChannelEI} report leakage currents on the order of $\sim$10 nA per MOSFET, which corresponds to a dissipated power of $6.5 \times 10^{-9}$ W per transistor under a 0.65 V supply voltage, which results in an energy cost several OOMs lower than that for a CMOS gate switch (see table \ref{tab:energy-costs}).
However, we opt against omitting this from the analysis for two reasons. 

The first reason is that while the energy costs per transistor are low, there can be many transistors dissipating all the time in a microprocessor, resulting in a large energy contribution. 
For instance, in a processor with 80 billion transistors (such as an NVIDIA H100 GPU, one of the premier data-center GPUs), the total power dissipation would be 520 W, which is roughly equal to the thermal design power of the H100 \cite{Andersch2022H100}.

The second reason is that there is some evidence that static power dissipation has grown quite quickly historically. 
For instance, \cite{Thompson1998MOSST, Nandyala2016ACT} point out that at the 1 $\mu$m node, leakage power dissipation had a contribution of about 0.01\% of the magnitude of dynamic power, while at the 100 nm node, this rose to roughly 10\%. 
ITRS projections from the early 2000s also predicted that static power dissipation would come to dominate overall processor power dissipation in the late 2000s \cite{Kim2003LeakageCM}, in part due to the large increase in the number of devices on ICs.
It appears that these forecasts have not panned out, e.g.  \cite{Wiltgen2013PowerStaticCMOS} found in 2013 that the majority of power dissipation is still dynamic rather than static, with \cite{Chen2014DianNaoAS} finding in 2014 that less than 2\% of the processor energy dissipation is due to leakage current.
However, it is unclear \textit{a priori} what this suggests about the limits to static power reductions.  

\section{The limits to energy efficiency}
\label{sec:energy-efficiency-limits}

Given the analysis in section \ref{sec:energy-cost-accounting}, we analyze the energy costs from three main sources: transistor switching, interconnect, and leakage power. 
For each energy source, we estimate lower bounds to the J/FLOP, and invert this to estimate the upper bound to the FLOP/J for CMOS-based technologies (see \ref{sec:CMOS-processors} for a definition).   

We consider processors able to perform operations at 4-bit precision in our analysis. Lower numerical precision requires less energy per FLOP and allows us to estimate an optimistic upper bound on efficiency. While common formats like FP32 or FP64 use higher precision, various emerging workloads are resilient to reduced precision \cite{Gupta2015DeepLW, Venkatesh2016AcceleratingDC}. For example, training deep neural networks can utilize precision as low as 8 or even 4 bits.\footnote{There has been a trend towards lower-precision formats for machine learning workloads \cite{Hobbhahn2023TrendsMLhardware}, and several proposals for 4-bit precision training of deep neural networks have recently been proposed \cite{Sun2020UltraLowP4}. Moreover, this could further be seen as a natural baseline for the level of precision since it is in line with existing estimates of the precision of operations performed by the human brain \cite{Lahiri2013AMF, Sandberg2008WholeBE, Bartol2015NanoconnectomicUB, carlsmith2020brain}.} By focusing on 4-bit operations, we establish an upper bound relevant to both current and future workloads that do not require full precision. In general, considering lower precision increases the estimated maximum FLOP/J by reducing the energy costs per operation.

Second, we consider processors that have a similar range of operating temperatures to modern chips, roughly between 273 K and 373 K (0$^\circ$C and 100$^\circ$C), with a representative temperature being 300 K. Processors often operate at mildly higher temperatures than this (possibly up to 373 K = 100$^\circ$C), but the rest of our arguments are sufficiently uncertain that assuming 300 K as a ballpark estimate for operating temperature introduces a negligible amount of additional error.

\subsection{Transistor switching}
\label{sec:transistor-switching}

The energy costs from transistor switching can be determined by starting from the Landauer limit, and adjusting it to account for reliability considerations.
We decompose our calculation as the product of two factors: 
$$\flopj = (Q_S \times N_T)^{-1},$$
where $Q_S$ is the heat dissipated per transistor switch, and $N_T$ is the number of transistors that need to be switched for each FLOP. 
We now estimate possible minimum values for each of $Q_S$ and $N_T$ to maximize the FLOP/J. 

\textbf{Dissipation per transistor switch $Q_S$}: 
At a minimum, storing a bit requires a potential energy barrier of $k_B T \ln 2$ based on the Landauer limit, which at 300 K gives $4 \times 10^{-21}$ J.

However, we also want to be able to store this information \emph{reliably}, and this likely requires an additional 1-2 orders of magnitude of energy to achieve, suggesting that transistors in CMOS processors could ideally operate with potential energy barriers on the order of $4 \times 10^{-20}$ J to $ 4 \times 10^{-19} $ J.

An important question is whether this potential energy stored in a transistor needs to be dissipated as heat. Current CMOS devices are \textit{irreversible}, which means they tend to discharge transistors rapidly and as a result lose most of the stored electrical potential energy to heat dissipation. However, adiabatic or reversible logic can, in principle, avoid these losses. In this paper, we explicitly consider discharging that occurs in the irreversible regime and assume that the processor does not employ adiabatic design to reduce the energy losses to heat dissipation. Using adiabatic methods requires devices to be substantially redesigned to avoid differences in propagation delays, so unless there is a paradigm change in the design of CMOS hardware this assumption should be safe. However, we want to explicitly note that it means reversible devices are out of the scope of our analysis.

The minimum energy cost of transistor switching in the irreversible regime has been analyzed by several other papers \cite{Agarwal2016EfficiencyLimitsLogicMemory, Frank2005ApproachingTP, Zhirnov2014MinimumEO}, with estimates ranging from around $3 \times 10^{-20}$ J/switch to around $6 \times 10^{-19}$ J/switch.
For our model we thus choose loose bounds of $[3 \times 10^{-20}, 10^{-18}]$ J/switch over this parameter. 

\textbf{Number of transistor switches per FLOP $N_T$}: 
We anchor our estimate of this parameter to existing processors that are highly optimized to maximize energy efficiency. 
In particular, \cite{Jayaprakasan2011EvaluationOT} describes a logic circuit for a 4-bit integer multiplier with 16 AND gates, 8 full adders, and 4 half adders.
With this layout, we then estimate that this circuit requires at a minimum of 368 transistors to implement, based on known limits on the number of transistors per gate.\footnote{For this calculation, we assume that we need a minimum of 4 transistors per OR gate, and 6 transistors per AND gate or XOR gate \cite{shashank2021why}. Given 16 AND gates (16*6 = 96 transistors), 8 full adders (8*28 = 224 transistors), and 4 half adders (4*12 = 48 transistors), we arrive at a total of 368 transistors.}
We reduce this to 300 transistors because circuits implemented as a whole rather than as a sum of parts can often achieve gains in transistor count, although these circuits are already heavily optimized and the gains are unlikely to be all that large.
Practical devices typically have an activity factor on the order of 10\% \cite{Alon2015EnergyEfficiencyLimitsCMOS}, so we estimate that roughly 30 transistors are switched per FLOP, at a minimum.

In practice, it is very difficult to design practical microprocessors that are so heavily optimized, and more transistor switches may need to be considered than just the transistors in a single multiplier unit. 
For instance, while our calculation assumed 6 transistors per AND gate, actual chips may need at least 1 OOM more to handle all of the control logic necessary to drive the computations as well as to improve reliability and speed.

Here is a real-world example of where we think such overheads make a significant difference. \cite{nvidia2023hopper} says that the NVIDIA H100 SXM has a clock frequency of 1.83 GHz and a total of 80 billion transistors. Given the stated dense FP16 performance of 989.4 TFLOP/s, if all transistors were dedicated to FP16 operations alone this would correspond to \( \approx 148,000 \) transistors per FP16 operation. Taking the H100's diverse capabilities at many different floating-point formats into account, it's likely that only a fraction of its total circuit complexity is oriented towards FP16 multiplications, but given the central role FP16 currently plays in machine learning applications we think this fraction should be at least around \( 1/3 \) or so, and possibly more than this.

If we naively assume quadratic scaling of the number of logic gates needed for an \( N \)-bit multiplication operation, assuming that something like the naive multiplication algorithm is implemented at the circuit level, then a 16-bit INT multiplier might cost around \( 16 \times 300 = 4800 \) transistors. If an FP16 multiplier has a similar cost, we're still looking at an overhead of around 1 OOM in a real-world GPU such as the H100 over the theoretical lower bound on the complexity of a multiplier.

Consequently, hardware specialization presents an important trade-off for the factors we consider in this section. For instance, as we specialize hardware in machine learning by moving from CPUs to GPUs to TPUs, we reduce overhead per operation by e.g. sharing control logic across many different arithmetic logic units (ALUs), but at the cost of making the hardware's use cases more limited. 

As we're thinking about potential upper bounds to energy efficiency, we will assume that the hardware we're concerned with is one that has low overhead per operation, though it's difficult to know in practice how much this overhead can be reduced past the rough 1 OOM estimate we compute above for the H100. We therefore adopt a relatively wide interval of \( [30, 3000] \) for the number of transistors that an optimal device will have to switch per 4-bit FLOP. The lower end of this corresponds to a low activity factor with very little overhead in implementation, while the high end corresponds to a 1 OOM control logic and other overheads and little savings coming from a low activity factor.

\textbf{FLOP/J}: Combining the uncertainty ranges for the parameters \( Q_S \) and \( N_T \) yields a range of $[3.3 \times 10^{14}, 1.1 \times 10^{18}] \flopj$ with a midpoint of \( 1.9 \times 10^{16} \flopj \).

\subsection{Interconnect}
\label{sec:flop-per-joule-bound}

As mentioned in section \ref{sec:energy-cost-accounting}, the energy contribution from interconnect is due to charging and discharging wire capacitances.
For a CMOS processor, we can determine this via the standard equation $E = \frac{1}{2} CV^2$, for capacitance $C$ and input voltage $V$ \cite{Miller2016AttojouleOF}.
Here $V$ is typically fixed exogenously, but $C$ can vary depending on the length of wire that needs to be charged up. 
We thus decompose the capacitance into the product of the capacitance per unit length $C_L$ and the total length of wire $L \times N$ that needs to be charged up, where $N$ is the number of wires charged up per FLOP and $L$ is the average length of each charged wire. 
Thus the relevant identity for determining the FLOP/J is given by
\begin{equation}
    \flopj = \left(C_L \times L \times N \times V^2\right) ^{-1}.
    \label{eq:flop-per-joule}
\end{equation}

\textbf{Capacitance per unit length $C_L$}:
This quantity tends to be very similar across a wide range of interconnects given that it does not change much with decreased scale \cite{Miller2016AttojouleOF}, and on-chip interconnects in present CMOS devices typically have $C_L \approx 2$ pF/cm \cite{Miller2009DeviceRF, Charkabarty2011Interconnects}. 
There are two primary ways of reducing $C_L$: (1) changing the geometry (e.g. increasing the interconnect spacing or decreasing the wire width), or (2) reducing the dielectric constant of the material surrounding the wires. 
However, these modifications are likely to be quite challenging, since reductions to the capacitances here are likely to lead to trade-offs with other figures of merit.

For instance, decreasing the wire width typically reduces the capacitance $C$ by a smaller factor than the factor by which it increases the resistance $R$, thus increasing signal delay $\tau = RC$.\footnote{To see why this is the case, consider a cuboidal wire with width $x$. When this is decreased, $R \propto \frac{1}{x}$ increases. However, $C \not\propto x$ in general—while this would be true for an idealized parallel plate capacitor, in real CMOS circuits there are fringe effects and interactions between different interconnect wires that result in $C$ scaling nonlinear with $x$. As a result, the effects on $R$ and $C$ do not cancel, and the $RC$ constant increases \cite{Abbas2020WiresAC}.}
Another example is that while reducing the dielectric constant $\kappa$ is an active area of research (i.e. ``low-$\kappa$ dielectric materials"), it is unlikely that the capacitance can be reduced by more than a factor of about 4$\times$.
This is because existing $\text{SiO}_2$ dielectrics have a dielectric constant of $\kappa \approx 4$, not to mention that low-$\kappa$ dielectrics are often too brittle for the harsh conditions of IC fabrication \cite{Shamiryan2004LowkDM}.\footnote{More radical approaches are possible—it is possible to get \textit{negative} dielectric constants (see e.g. \cite{Fuhrer2021NegativeCapacitance, Datta2022TowardAS}), but we consider these technologies to be out of scope. For instance, \cite{Fuhrer2021NegativeCapacitance} describes transistors where electron spins become important, and \cite{Datta2022TowardAS} suggests the use of ferroelectric dielectrics which help reduce energy loss through a fundamentally different physical mechanism to conventional approaches.} 

A third approach to reduce \( C_L \) is to increase the spacing between different layers of the processor and between different wires on the same layer. As capacitance falls off with distance, this might seem like an obvious solution. However, practical devices are limited in size if they require fine control over the computations or other kinds of high-frequency data movement. As an example, at the H100's clock frequency of 1.83 GHz, a light signal in a vacuum can only travel a distance of 16 centimeters per clock cycle. So trying to reduce capacitance per unit length by making devices less dense comes at the cost of having to make computations increasingly more localized, making this an impractical choice for trying to lower \( C_L \).

Given the above considerations, and that $C_L$ has not been meaningfully reduced over the last 1-2 decades\footnotemark, it appears difficult to yield major improvements in the capacitance without fairly radical changes in the technologies used. 
\footnotetext{We gather data support this claim based on the ITRS and IRDS reports between 2007 and 2022—over this entire period we find that $C_L$ remains at around 2 pF/cm. \href{https://docs.google.com/spreadsheets/d/1hM-XCJ2Yob-Ne1hHv9164Px8spMxn-HcT5Fj8qSrx5E/edit\#gid=272816530}{[Data]}.}
This is in line with the IRDS 2022 forecasts that $C_L$ will not decrease over the next 15 years \cite{IEEE2022IRDSmoreMoore}. 
As a lower bound, we assume that optimistically $C_L$ can be improved by roughly a factor of 10$\times$ compared to today, such that our bounds for this parameter are thus $[2 \times 10^{-11}, 2 \times 10^{-10}]$ F/m.

\textbf{Average wire length $L$}: 
While each FLOP is performed in a Floating Point Unit (FPU), in actual microprocessors the FLOP/J will also depend on the energy costs associated with memory accesses (e.g. to external DRAM), involving longer wires than in the FPU itself.

To obtain a lower bound to $L$ (for an upper bound to the FLOP/J) we primarily consider just the wires in the FPU, conservatively supposing this to be representative. We then estimate $L$ using a bottom-up argument from lower bounds to transistor lengths.

Our approach starts from projections of the minimum width of future transistors, and bases our estimates of average interconnect length in terms of the number of such transistor widths.
Existing estimates of minimum transistor gate widths typically range from 0.5 nm \cite{Simmons2012singleAtomTransistor}\footnote{The lattice constant of a Silicon with a diamond cubic crystal lattice structure \cite{Tiesinga2021CODATARV}.} to about $\sim$3 nm \cite{Muller1999TheES, Nawrocki2010PhysicalLF}.
Transistor widths are typically on the order of 5$\times$ of gate widths \cite{Schor2019tsmc}, and if this relationship continues to hold, the minimum transistor width would be around 2.5—15 nm.

To make inferences about how interconnect length relates to transistor dimensions, we can look at existing hardware. Currently, the average area per transistor on contemporary GPU chips appears to be around $ 10^4 \nm^2 $. For example, \cite{Andersch2022H100} states that the die area of an H100 SXM GPU is \( 814 \mm^2 \) and the GPU has a total of 80 billion transistors. Dividing, we get an area of \( 10,175 \nm^2 \) per transistor, suggesting a transistor spacing length of \( \approx 100 \nm \) on the die. This is to be contrasted with the gate pitch of around \( 50 \nm \) used by state-of-the-art processes such as the 3 nm process \cite{IEEE2022IRDSmoreMoore}. The \( 2 \times \) discrepancy could be due to several factors: not all of the available area of the die used for packing transistors as densely as possible, NVIDIA's "transistors" being more complex than the ones considered in \cite{IEEE2022IRDSmoreMoore}, \textit{et cetera}.

On the other hand, we can use the formula \( (C_L \times L \times N \times V)^{-1} \) for the interconnect heat dissipation losses to try to infer \( L \) from the other known parameters, at least in an approximate way. We know \( C_L \approx 2 \) pF/cm, \( N \approx 8 \times 10^{11} \) transistors and \( V \approx 1 \) volt for the H100. \cite{nvidia2023hopper} also says that the TDP of an H100 SXM is \( 700 \watt \). Assuming an activity factor of \( \alpha \), so that a fraction \( \alpha \) of all transistors are switched per clock cycle, and solving the equation

\[ \alpha \times C_L \times L \times N \times V = \frac{700 \watt}{1.83 \, \text{GHz}} \]

for \( L \) yields \( L \approx (24 \nm)/\alpha \) per transistor. A typical activity factor on the order of \( 10 \% \) \cite{Alon2015EnergyEfficiencyLimitsCMOS} means \( L \approx 240 \nm \), which is roughly in line with the transistor spacing length of \( \sim 100 \nm \) that was calculated above and \( 5 \) times the \( 3 \nm \) node gate pitch of \( 48 \nm \). If this rough proportionality is assumed to hold when devices are miniaturized further, we could assume that interconnect lengths per transistor will scale proportionally with the gate pitch with a constant of around \( 5 \). The IRDS roadmap \cite{IEEE2022IRDSmoreMoore} is pessimistic about progress on this dimension, projecting that the gate pitch of \( 48 \nm \) at the \( 3 \nm \) node will merely fall to \( 38 \nm \) by 2037. However, more fundamental arguments suggest smaller transistors are possible.

Given that interconnect lengths per transistor at the \( 3-5 \nm \) node with a gate pitch of \( 48 \nm \) seem to be around \( 240 \nm \), we take the \( 38 \nm \) estimate of \cite{IEEE2022IRDSmoreMoore} as an upper bound of what's possible and take \( 2.5 \nm \) as a conservative lower bound. Multiplying by a factor of \( 5 \), this gives us an interval \( [12.5 \nm, 190 \nm] \) for \( L \).

\textbf{Number of wires $N$ charged per FLOP}: 
We estimate this based on the number of transistors that need to be switched per FLOP and assume that at least one wire needs to be charged per transistor switch. 
We thus follow the estimate for transistor switches $N_T$ and arrive at a range of $[30, 3000]$ wires for $N$. Note that this estimate already incorporates the possibility that the activity factor might be as low as \( 10 \% \), so we don't separately include the activity factor in our final calculation.

\textbf{Supply voltage $V$}: 
CMOS processors can potentially be run at significantly below the roughly 0.7 V of today \cite{IEEE2022IRDSexec}. 
For instance, Swanson and Meindl derive a theoretical lower bound of 36 mV for the supply voltage, such that MOSFETs can still be switched \cite{Meindl2000TheFL, Swanson1972IonimplantedCM}.
In particular, the minimum operational voltage $V_\text{min}$ is given by \cite{Zhai2005TheLO}: 
$$V_\text{min} = 2 V_T \left( 1 + \frac{S_S}{\ln 10 \cdot V_T}\right),$$
where $V_T = \frac{k_B T}{e}$ is the thermal voltage given Boltzmann's constant $k_B$, temperature $T$ and fundamental charge $e$.

$S_S$ is the subthreshold swing, and Swanson and Meindl choose $S_S = 60$ mV/decade in the ideal case to obtain 36 mV (at 300 K). 

However, Zhai \textit{et al.} (2005) argue that this bound for the operational frequency is however not energy-optimal, due to increased leakage energy from voltage scaling \cite{Zhai2005TheLO}. 
They instead find that a minimum at around 0.2 V, e.g. for a $16 \times 16$ multiplier circuit. 
This gives similar results from Zhai \textit{et al.} (2004) for a variety of different circuits \cite{Zhai2004TheoreticalAP}. 

Furthermore, \cite{Datta2022TowardAS} argue that the operating voltage needs to be at least 0.5 V in practice, with 0.2 V needed to keep leakage currents acceptably low and an additional 0.3 V to deliver sufficient current. 

These estimates however depend on how well optimized processors are to handle static power dissipation. 
If future FETs admit significantly smaller leakage currents, the balance at which static and dynamic power dissipation could shift, making lower values of $V$ possible.

We think the estimates from \cite{Meindl2000TheFL, Swanson1972IonimplantedCM} are theoretically sound. However, outside of a laboratory setting and in a practical device, it seems exceedingly difficult to make complex logic work with just $ 36 $ mV of voltage given the ideal subthreshold swing of $ 60 $ mV/decade. In practice, the error correction required to make logic nodes work correctly with at most \( 10^{36/60} \approx 4 \times \) current differences between the subthreshold and threshold regimes of a transistor would present large overheads that would most likely eliminate the gains of scaling voltages down to such a level. 

To reconcile the estimates that look most realistic to us, we take \( 0.1 \) volts as our lower bound and the projection of $ 0.6 $ volts from \cite{IEEE2022IRDSmoreMoore} as the upper bound of our uncertainty interval over the supply voltage. The choice of \( 0.1 \) volts as the lower bound is based on wanting the number to be low enough to include the results from \cite{Zhai2004TheoreticalAP, Zhai2005TheLO} while high enough to exclude values such as $ 36 $ mV that we regard as unrealistic for practical devices.

\textbf{FLOP/J}:
If we combine all of the above, we arrive at a range of $[1.5 \times 10^{13}, 1.3 \times 10^{18}] \flopj$, though importantly the distribution we have for the maximum attainable energy efficiency considering only interconnect losses is \textit{not uniform} over this interval. 

\subsection{Leakage power}
The final part of our analysis in this section pertains to the leakage power.
As mentioned in section \ref{sec:energy-cost-accounting}, this has shown some signs of becoming increasingly important over time—the goal of this section is thus to verify whether or not these costs need to be considered in our final model for the FLOP/J. 

While it is the case that static power was quickly becoming important in the 2000s, a crucial counter consideration was the the increased popularity of FinFETs in the 2010s \cite{Kamal2022TheSA}.
These transistors have orders of magnitude lower leakage currents than preceding MOSFETs \cite{Verma2020ProcessVA}, drastically decreasing static energy costs.  
For instance, \cite{Badran2019LowLC} finds FinFET leakage currents on the order of 20 pA/$\mu$m at a roughly 10 nm node, which corresponds to $2 \times 10^{-13}$ A of leakage current per transistor. 
This would correspond to a total leakage power of $1.04 \times 10^{-2}$ W (again assuming 0.65 V), four OOMs lower than the example with MOSFETs we considered in section \ref{sec:energy-cost-accounting}. 

To determine the extent to which leakage power can be reduced, we turn to estimates of minimum leakage current. 
To this end, \cite{Huang1999Sub5F} estimates that FinFET leakage currents may be able to reach $10^{-12}$ A/$\mu$m.
Given our previously mentioned estimates of 0.5 nm—3 nm minimum gate lengths, this corresponds to a current of $5 \times 10^{-16}$ A to $3 \times 10^{-15}$ A. 
Combined with our estimate of 0.1 V for the supply voltage, this corresponds to $5 \times 10^{-17}$ W to $3 \times 10^{-16}$ W per transistor, or $4 \times 10^{-6}$ W to $2.4 \times 10^{-5}$ W in a processor with 80 billion transistors. 

We can compare this with the dynamic energy costs from transistor switching and interconnects, which had roughly $10^{18}$ and $10^{19}$ FLOP/J respectively \emph{in the most optimistic estimates}.
Since the H100 has a performance of around $10^{14}$ FLOP per second, the corresponding wattage due to dynamic dissipation is $10^{14} / 10^{19} = 10^{-5}$ W, which is just comparable with the calculations for static power. 
However the effects from static power are much smaller in most other cases—e.g. a bound of $10^{16}$ FLOP/J yields $10^{-2}$ W given H100 performance, which is 3 to 4 OOMs higher than the contribution from static power. 
Even pessimistically assuming that leakage currents do not decrease from the $10^{-13}$ A of today, this only brings us to $10^{-4}$ W to $10^{-3}$ W, which is 1 to 2 OOMs lower than the best guess scenario. 
These calculations suggest that for our purposes of deriving an upper bound to the FLOP/J, static power contributions are likely to be negligible, at least as long as supply voltages are kept sufficiently high. 

\section{Model}
We can now combine transistor and interconnect models into a complete model predicting the maximum achievable FLOP/J of an optimized CMOS processor. We've established that the J/FLOP comes primarily from two main sources, namely switching capacitances in transistors and wire interconnects. 
Thus we write $$E = E_\text{transistor} + E_\text{interconnect}.$$
Here $E$ is the total energy dissipated per FLOP, $E_\text{transistor}$ is the contribution from transistor switching and $E_\text{interconnect}$ the contribution from interconnects. 
We also have that
$$E_\text{transistor} = Q_S \times N_T,$$
$$E_\text{interconnect} = C_L \times L \times N \times V^2.$$
The FLOP/J is then determined by taking the reciprocal of $E$, and we can obtain a probability distribution over this via Monte Carlo simulation.
To estimate the distribution of possible FLOP/J values, we perform Monte Carlo sampling across plausible ranges for each parameter. The sampling uses log-uniform distributions, reflecting uncertainty in parameter values.

Specifically, the log-scale parameter bounds are transformed to [Lower, Upper] intervals. Random sampling on the log-transformed intervals provides a weakly informative prior distribution. We think our uncertainty over the number of transistors required per FLOP should be independent of our uncertainty over other parameters, but it's possible there are some complex dependencies between \( C_L, L \), and \( V \) that we have neglected so far.

To take this into account in a crude way instead of neglecting it altogether, we use a multivariate Gaussian copula with different parameters having a constant correlation of \( 0 < \rho < 1 \), and assume \( \rho \) is distributed as \( \sim |2X - 1| \) where \( X \) follows a \( \text{Beta}(2, 2) \) distribution. We then reverse the map taking the latent Gaussian variables to parameter values with probability \( 1/2 \) during Monte Carlo sampling to take into account that we don't know the direction of the correlations between parameters.\footnote{ The code for the exact process we use can be found in the accompanying \href{https://colab.research.google.com/drive/18WKCe6QrZ8qCA5v0pZGaNhq7ovCXmqFL?usp=sharing}{Colab notebook}.}

\begin{table}[h]
\centering
\renewcommand{\arraystretch}{1.15} 
\begin{tabular}{c|c}
\textbf{Variable} & \textbf{Range} \\
\hline
Energy per switch $Q_S$ & $[3 \times 10^{-20}, 10^{-18}]$ J/switch \\
Number of switches $N_T$ & [30, 3000] switches\\
$1/E_\text{transistor}$ & $[3.3 \times 10^{14}, 1.1 \times 10^{18}]$ FLOP/J \\
\hline
Capacitance per unit length $C_L$ & $[2 \times 10^{-11}, 2 \times 10^{-10}]$ F/m \\
Avg. wire length $L$ & $[12.5 \times 10^{-9}, 190 \times 10^{-9}]$ m \\
Number of wires $N$ & $[30, 3000]$ wires \\
Supply voltage $V$ & $[0.1, 0.6]$ V \\
$1/E_\text{interconnect}$ & $[1.5 \times 10^{13}, 1.3 \times 10^{18}]$ FLOP/J \\
\end{tabular}
\caption{Summary of all upper and lower bounds for the key variables in the described model.}
\label{tab:estimates}
\end{table}

\begin{figure}
    \centering
    \includegraphics[width=0.5\textwidth]{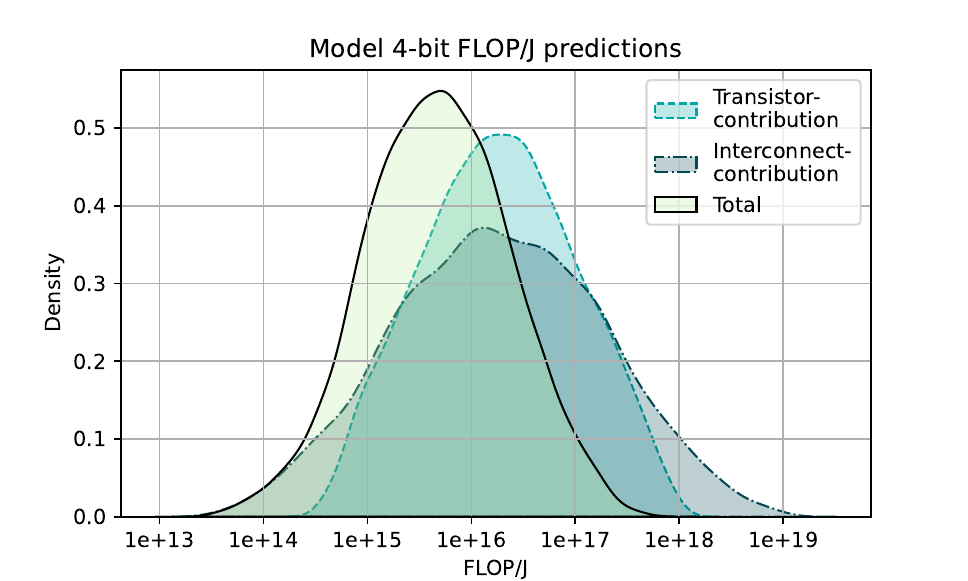}
\caption{Density distribution of predicted maximum FLOP/J (FLOP per Joule). The plot includes transistor-based, interconnect-based, and total, derived using log-uniform sampling across model parameters. The total represents the sum of the transistor and interconnect model predictions, accounting for both components of power consumption.}
    \label{fig:flopj}
\end{figure}

The effect of this modification is to widen the energy efficiency distribution coming from the interconnect method. Despite its ad-hoc nature, we believe not taking possible correlations into account at all makes our estimate overconfident, and even a rough method to take unknown correlations into account should be better than assuming all variables are jointly independently distributed by default.

Figure \ref{fig:flopj} shows the distribution we obtain using this method for FLOP/J taking both interconnect and transistor switching losses into account.
The geometric mean of this distribution is around $ 4.7 \times 10^{15}$ FLOP/J, with a standard deviation (in log space) of around 0.7 OOMs.

\section{Discussion}

\subsection{Implications}
\label{sec:implications}

The current energy efficiency of state-of-the-art GPUs such as the H100 \cite{nvidia2023hopper} for dense floating point operations is around $ (10^{15} \, \, \text{FP16/s})/(700 \watt) \approx 1.4 \times 10^{12} $ FP16/J. Given that we expect ideal energy costs to scale quadratically with precision due to the quadratic complexity of the naive multiplication algorithm which is optimal at small precisions, our geometric mean forecast of $ 4.7 \times 10^{15} $ FP4/J corresponds to a forecast of $ 4.7 \times 10^{15}/16 = 2.9 \times 10^{14} $ FP16/J. Our results, therefore, suggest that there is a \( 50 \% \) chance that further improvements in energy efficiency will cease after another \( 290/1.4 \approx 207\)-fold of improvement on existing technology.

An important implication is economic in nature: so long as power remains expensive, our results set an upper bound on how many floating point operations can be purchased on a fixed budget, regardless of how cheap hardware manufacturing itself becomes.

Today, the cost to end users of specialized hardware such as the H100 is dominated almost entirely by the price of the hardware: \cite{eadline2023nvidia} reports that the H100 was selling for \$30,000 apiece in August 2023. The useful mean lifetime of a GPU is likely on the order of 5 years, both because of depreciation and because of newer hardware making older hardware obsolete, and \cite{eia2023power} states that industrial customers in the US have to pay on the order of \( 10 \) cents per kWh of power in 2023. At these rates, running the H100 at a TDP of \( 700 \watt \) for 5 years costs around \$3,000—only \( 10 \% \) of the cost of the hardware itself.

So far power costs have been a small fraction of overall expenses on specialized computing applications. However, even if hardware continues to get cheaper on a nominal FLOP/s/\$ basis, our results put an upper bound on just how cheap computing can become as long as energy prices remain flat. This has significant implications for many domains: for example, an application that motivated the research leading to this paper is the training of large machine learning systems, where an end to FLOP/\$ scaling can make AI training runs beyond some scale infeasible.

All of these implications are conditional on the current hardware paradigm not being abandoned in favor of one that would violate some key assumptions of our analysis in this paper. We discuss these assumptions in more detail in \hyperref[sec:CMOS-processors]{Appendix A}, but it's worth highlighting the most important one of them once more: our calculations only hold in the irreversible regime where energy stored in transistors and stray capacitance is almost entirely lost as heat dissipation whenever these capacitors are switched. Assessing the feasibility of adiabatic computing methods is beyond the scope of this paper and we direct the interested reader to \cite{Frank2020ReversibleCW}.

\subsection{Further work}

Given that energy prices have not come down anywhere near as quickly as hardware prices, and that power costs of running specialized computing hardware are becoming comparable to marginal manufacturing costs before producer markups, the question of studying energy efficiency in computing gains importance with the passage of time. Far from being the final say on the subject, we think this paper should be viewed as a first-pass attempt at estimating limits to FLOP/J scaling that future work will substantially improve on.

There are many ways in which the crude analysis in this paper may be improved. A more detailed understanding of interconnect dimensions, the feasibility of aggressive scaling down of supply voltages, and a more rigorous accounting of the correlations between all the variables involved in the analysis would lead to more accurate predictions. The models themselves can also be refined, e.g. by properly taking static power dissipation into account in the small voltage regime or in other miscellaneous ways.

\section{Conclusion}
In conclusion, this paper presents an approach to estimating the maximum energy efficiency in FLOP/J of CMOS processors. We first performed an accounting of primary energy costs, which identified two main sources of dissipation: (1) switching transistors and (2) switching wire capacitances. This analysis was then used to establish a model that predicts a distribution over the maximum FLOP/J, where interval estimates were established for each individual model parameter. 

The model has a geometric mean estimate of $4.7 \times 10^{15}$ FLOP/J as the maximum energy efficiency for CMOS devices, with a log-space standard deviation of around 0.7 OOMs in FLOP/J. Compared to current state-of-the-art graphics cards such as the H100, this represents an improvement of roughly 2.3 OOMs, or about 207-fold.

This approach opens up the possibility for critique and improvement of the model assumptions, a more transparent, verifiable, and iterative approach to forecasting the fundamental limits of CMOS processors.

\textbf{Acknowledgements}. We would like to thank Christopher Phenicie, Jaime Sevilla, Lennart Heim, Fabian Peddinghaus, Maxwell Anderson, Avinoam Kolodny, Anders Sandberg, Michael P Frank, Tom Davidson, Paul Cruickshank, and Paul Christiano for helpful discussions and feedback, as well as  \href{https://electronics.stackexchange.com/questions/669375/what-is-the-average-length-of-interconnect-wire-that-needs-to-be-charged-dischar}{various users of Electrical Engineering StackExchange}.

\section*{Appendix}

\subsection*{Appendix A: Characterizing CMOS processors}
\label{sec:CMOS-processors}

In order for our estimates of the upper bound to be sufficiently precise, we need to define ``CMOS processors" with appropriate specificity—this determines what technologies this model's prediction is intended to apply to, and which kinds of technological advancements are included or excluded in our analysis. 
The definition we provide aims to capture what is commonly meant by the ``present hardware paradigm" and allows us to derive tractable bounds using standard energy analysis techniques, while still allowing some degree of leeway for future energy efficiency improvements. 

Roughly, we want our definition to refer to technologies listed in the ``More Moore" category and not the ``Beyond CMOS" category of the 2022 IRDS reports \cite{IEEE2022IRDSexec}, which are standard primary references for developments in the computing hardware industry.
As such, ``CMOS processors" are defined as processors with the following criteria: 
\begin{itemize}
    \item \textbf{Logic operations are performed digitally}, thus excluding analog technologies e.g. floating gate transistors. 
    \item \textbf{Computational states are based on electron charge}, which excludes technologies such as optical interconnects and spin-based computational states.
    \item \textbf{Operations are mostly performed irreversibly} and thus are subject to the Landauer limit. 
    Approaches such as adiabatic switching are possible for achieving much lower energy dissipation, but such technologies tend to be very challenging to implement in digital CMOS \cite{Frank2020ReversibleCW} and are largely out of scope for this paper.
    \item \textbf{Processors are based on a Von Neumann architecture}, where memory and logic are separated via a bus, hence excluding in-memory computing technologies, e.g. memristors.
    \item \textbf{No engineered nanomaterials}\footnotemark, which excludes technologies such as carbon nanotube interconnects but not doped semiconductors, since the latter does not have a regular structure at nanometer scales.
    \footnotetext{We adopt the definition of ``nanomaterials" from the International Organization for Standardization, namely ``material with any external dimension in the nanoscale or having an internal structure or surface structure in the nanoscale [1 nm to 100 nm]" \cite{ISO2015nanotechnologies}.}
\end{itemize}

This specification importantly differs from ``traditional" CMOS in that we allow certain technological integrations, such as non-planar FinFETs \cite{Maszara2013FinFETsT}, whereas ``traditional" CMOS devices are strictly speaking restricted to planar MOSFETs. 
As such, our definition applies to the vast majority of microprocessors that are available at the time of writing, including those at the state of the art (e.g. NVIDIA H100s)—this is crucial for ensuring the practical relevance of our estimates.

\printbibliography

@article{Miller2016AttojouleOF,
  title={Attojoule Optoelectronics for Low-Energy Information Processing and Communications},
  author={David A. B. Miller},
  journal={Journal of Lightwave Technology},
  year={2016},
  volume={35},
  pages={346-396}
}

@article{Miller2009DeviceRF,
  title={Device Requirements for Optical Interconnects to Silicon Chips},
  author={David A. B. Miller},
  journal={Proceedings of the IEEE},
  year={2009},
  volume={97},
  pages={1166-1185}
}

@misc{Charkabarty2011Interconnects,
    author = {Krishnendu Chakrabarty},
    title = {ECE 261: CMOS VLSI Design Methodologies - Interconnects},
    year = {2011},
    url = {http://people.ee.duke.edu/~krish/teaching/Lectures/AdvancedTopicsInterconnect.pdf}
}

@report{IEEE2022IRDSexec,
    author = {IEEE},
    title = {International Roadmap for Devices and Systems 2022 Edition -- Executive Summary},
    year = {2022},
    url = {https://irds.ieee.org/images/files/pdf/2022/2022IRDS_ES.pdf}
    }

@report{IEEE2022IRDSmoreMoore,
    author={IEEE},
    title={International Roadmap for Devices and Systems 2022 Edition -- More Moore},
    year={2022},
    url={https://irds.ieee.org/images/files/pdf/2022/2022IRDS_MM.pdf}}

@article{Koomey2011ImplicationsOH,
  title={Implications of Historical Trends in the Electrical Efficiency of Computing},
  author={Jonathan G. Koomey and Stephen Berard and Marla Sanchez and Henry Wong},
  journal={IEEE Annals of the History of Computing},
  year={2011},
  volume={33},
  pages={46-54}
}

@article{Maszara2013FinFETsT,
  title={FinFETs — Technology and circuit design challenges},
  author={Witold P. Maszara and M.-R. Lin},
  journal={2013 Proceedings of the European Solid-State Device Research Conference (ESSDERC)},
  year={2013},
  pages={3-8}
}

@incollection{Ng22powerCMOS,
author = {Len Luet Ng and Kim Ho Yeap and Magdalene Wan Ching Goh and Veerendra Dakulagi},
title = {Power Consumption in CMOS Circuits},
booktitle = {Electromagnetic Field in Advancing Science and Technology},
publisher = {IntechOpen},
address = {Rijeka},
year = {2022},
editor = {Hai-Zhi Song and Kim Ho Yeap and Magdalene Wan Ching Goh},
chapter = {5},
doi = {10.5772/intechopen.105717},
url = {https://doi.org/10.5772/intechopen.105717}
}

@article{Wiltgen2013PowerStaticCMOS,
  title={Power consumption analysis in static CMOS gates},
  author={Alberto Wiltgen and Kim A. Escobar and Andr{\'e} In{\'a}cio Reis and Renato Perez Ribas},
  journal={2013 26th Symposium on Integrated Circuits and Systems Design (SBCCI)},
  year={2013},
  pages={1-6}
}

@article{Ranurez2006gateTunnelingCurrentMOS,
  title={A review of gate tunneling current in MOS devices},
  author={Juan C. Ranu{\'a}rez and M. Jamal Deen and Chih-Hung Chen},
  journal={Microelectron. Reliab.},
  year={2006},
  volume={46},
  pages={1939-1956}
}

@article{Anderson2023OpticalT,
  title={Optical Transformers},
  author={Maxwell G. Anderson and Shifan Ma and Tianyu Wang and Logan G. Wright and Peter Leonard McMahon},
  journal={ArXiv},
  year={2023},
  volume={abs/2302.10360}
}

@article{Frank2005ApproachingTP,
  title={Approaching the physical limits of computing},
  author={Michael P. Frank},
  journal={35th International Symposium on Multiple-Valued Logic (ISMVL'05)},
  year={2005},
  pages={168-185}
}

@inproceedings{Magen2004InterconnectpowerDI,
  title={Interconnect-power dissipation in a microprocessor},
  author={Nir Magen and Avinoam Kolodny and Uri C. Weiser and Nachum Shamir},
  booktitle={International Workshop on System-Level Interconnect Prediction},
  year={2004}
}

@article{Koomey2015energyEfficiencyUpdate,
    title={Moore's law might be slowing down, but not energy efficiency},
    author={Jonathan Koomey and Samuel Naffziger},
    journal={IEEE Spectrum},
    year={2015},
    url={https://spectrum.ieee.org/moores-law-might-be-slowing-down-but-not-energy-efficiency}
}

@article{Hobbhahn2023TrendsMLhardware,
    title={Trends in Machine Learning Hardware (Forthcoming)},
    author={Marius Hobbhahn and Lennart Heim and Gökçe Aydos},
    year={2023}
}

@article{Agarwal2016EfficiencyLimitsLogicMemory,
  title={Energy efficiency limits of logic and memory},
  author={Sapan Agarwal and Jeanine E. Cook and Erik P. Debenedictis and Michael P. Frank and Gert Cauwenberghs and Sriseshan Srikanth and Bobin Deng and Eric R. Hein and Paul G. Rabbat and Thomas M. Conte},
  journal={2016 IEEE International Conference on Rebooting Computing (ICRC)},
  year={2016},
  pages={1-8}
}

@article{Landauer1961IrreversibilityAH,
  title={Irreversibility and heat generation in the computing process},
  author={Rolf Landauer},
  journal={IBM J. Res. Dev.},
  year={1961},
  volume={5},
  pages={183-191}
}

@article{Manipatruni2018ScalableEM,
  title={Scalable energy-efficient magnetoelectric spin–orbit logic},
  author={Sasikanth Manipatruni and Dmitri E. Nikonov and Chia-Ching Lin and Tanay A. Gosavi and Huichu Liu and Bhagwati Prasad and Yen-Lin Huang and Everton Bonturim and Ramamoorthy Ramesh and Ian A. Young},
  journal={Nature},
  year={2018},
  volume={565},
  pages={35-42}
}

@article{Nikonov2015BenchmarkingOB,
  title={Benchmarking of Beyond-CMOS Exploratory Devices for Logic Integrated Circuits},
  author={Dmitri E. Nikonov and Ian A. Young},
  journal={IEEE Journal on Exploratory Solid-State Computational Devices and Circuits},
  year={2015},
  volume={1},
  pages={3-11}
}

@inproceedings{Alon2015EnergyEfficiencyLimitsCMOS,
  title={Energy efficiency limits of digital circuits based on CMOS transistors},
  author={Elad Alon and Tsu-Jae King Liu and Kelin J. Kuhn},
  year={2015}
}

@article{Sze2017EfficientProcessingDNNs,
  title={Efficient Processing of Deep Neural Networks: A Tutorial and Survey},
  author={Vivienne Sze and Yu-hsin Chen and Tien-Ju Yang and Joel S. Emer},
  journal={Proceedings of the IEEE},
  year={2017},
  volume={105},
  pages={2295-2329}
}

@article{Horowitz2014ComputingsEnergyProblem,
  title={1.1 Computing's energy problem (and what we can do about it)},
  author={Mark Horowitz},
  journal={2014 IEEE International Solid-State Circuits Conference Digest of Technical Papers (ISSCC)},
  year={2014},
  pages={10-14}
}

@article{Vogelsang2010UnderstandingDRAMenergy,
  title={Understanding the Energy Consumption of Dynamic Random Access Memories},
  author={Thomas Vogelsang},
  journal={2010 43rd Annual IEEE/ACM International Symposium on Microarchitecture},
  year={2010},
  pages={363-374}
}

@article{Keckler2011GPUsAT,
  title={GPUs and the Future of Parallel Computing},
  author={Stephen W. Keckler and William J. Dally and Brucek Khailany and Michael Garland and David Glasco},
  journal={IEEE Micro},
  year={2011},
  volume={31},
  pages={7-17}
}

@article{Fuhrer2021NegativeCapacitance,
  title={Proposal for a Negative Capacitance Topological Quantum Field-Effect Transistor},
  author={Michael S. Fuhrer and Mark T. Edmonds and Dimitrie Culcer and Mustansar Nadeem and X L Wang and Nikhil V. Medhekar and Yuefeng Yin and Jared H. Cole},
  journal={2021 IEEE International Electron Devices Meeting (IEDM)},
  year={2021},
  pages={38.2.1-38.2.4}
}

@inproceedings{Moiseev2015AnOverviewVLSIinterconnect,
  title={An Overview of the VLSI Interconnect Problem},
  author={Konstantin Moiseev and Avinoam Kolodny and Shmuel Wimer},
  year={2015}
}

@article{Shamiryan2004LowkDM,
  title={Low-k dielectric materials},
  author={D. Shamiryan and Thomas Abell and Francesca Iacopi and Karen Maex},
  journal={Materials Today},
  year={2004},
  volume={7},
  pages={34-39}
}

@inproceedings{Sun2020UltraLowP4,
  title={Ultra-Low Precision 4-bit Training of Deep Neural Networks},
  author={Xiao Sun and Naigang Wang and Chia-Yu Chen and Jiamin Ni and Ankur Agrawal and Xiaodong Cui and Swagath Venkataramani and Kaoutar El Maghraoui and Vijayalakshmi Srinivasan and K. Gopalakrishnan},
  booktitle={Neural Information Processing Systems},
  year={2020}
}

@article{Simmons2012singleAtomTransistor,
  title={A single atom transistor},
  author={Michelle Yvonne Simmons},
  journal={2012 IEEE Silicon Nanoelectronics Workshop (SNW)},
  year={2012},
  pages={1-1}
}

@article{Tiesinga2021CODATARV,
  title={CODATA Recommended Values of the Fundamental Physical Constants: 2018.},
  author={Eite Tiesinga and Peter J. Mohr and David B. Newell and Barry N. Taylor},
  journal={Journal of physical and chemical reference data},
  year={2021},
  volume={93 2}
}

@article{Muller1999TheES,
  title={The electronic structure at the atomic scale of ultrathin gate oxides},
  author={David A. Muller and Thomas W. Sorsch and S. Moccio and Frieder H. Baumann and Kenneth Evans-Lutterodt and Gregory Timp},
  journal={Nature},
  year={1999},
  volume={399},
  pages={758-761}
}

@article{Nawrocki2010PhysicalLF,
  title={Physical limits for scaling of integrated circuits},
  author={Waldemar Nawrocki},
  journal={Journal of Physics: Conference Series},
  year={2010},
  volume={248},
  pages={012059}
}

@article{Kim2003LeakageCM,
  title={Leakage Current: Moore's Law Meets Static Power},
  author={Nam Sung Kim and Todd M. Austin and David Blaauw and Trevor N. Mudge and Kriszti{\'a}n Flautner and Jie S. Hu and Mary Jane Irwin and Mahmut T. Kandemir and Narayanan Vijaykrishnan},
  journal={Computer},
  year={2003},
  volume={36},
  pages={68-75}
}

@article{Datta2022TowardAS,
  title={Toward attojoule switching energy in logic transistors},
  author={Suman Datta and Wriddhi Chakraborty and Marko Radosavljevic},
  journal={Science},
  year={2022},
  volume={378},
  pages={733 - 740}
}

@article{Nandyala2016ACT,
  title={A circuit technique for leakage power reduction in CMOS VLSI circuits},
  author={Venkata Ramakrishna Nandyala and Kamala Kanta Mahapatra},
  journal={2016 International Conference on VLSI Systems, Architectures, Technology and Applications (VLSI-SATA)},
  year={2016},
  pages={1-5}
}

@inproceedings{Thompson1998MOSST,
  title={MOS Scaling: Transistor Challenges for the 21st Century},
  author={Scott E. Thompson},
  year={1998}
}

@article{Rjoub2020AccurateLC,
  title={Accurate leakage current models for MOSFET nanoscale devices},
  author={Abdoul Rjoub and Mamoun Mistarihi and Nedal Al Taradeh},
  journal={International Journal of Electrical and Computer Engineering (IJECE)},
  year={2020}
}

@article{Stellari2006SwitchingTE,
  title={Switching Time Extraction of CMOS Gates using Time-Resolved Emission (TRE)},
  author={Franco Stellari and Alberto Tosi and Peilin Song},
  journal={2006 IEEE International Reliability Physics Symposium Proceedings},
  year={2006},
  pages={566-573}
}

@article{Andersch2022H100,
    title={NVIDIA Hopper Architecture In-Depth},
    author={Michael Andersch and Greg Palmer and Ronny Krashinsky and Nick Stam and Vishal Mehta and Gonzalo Brito and Sridhar Ramaswamy},
    year={2022},
    url={https://developer.nvidia.com/blog/nvidia-hopper-architecture-in-depth/}}

@article{Kamal2022TheSA,
  title={The Silicon Age: Trends in Semiconductor Devices Industry},
  author={Kamal Y. Kamal},
  journal={Journal of Engineering Science and Technology Review},
  year={2022}
}

@article{Verma2020ProcessVA,
  title={Process variation and analysis of FinFET for low-power applications},
  author={Shekhar Verma and Suman Lata Tripathi},
  journal={IOP Conference Series: Materials Science and Engineering},
  year={2020},
  volume={872}
}

@misc{Schor2019tsmc,
    title={TSMC 7nm HD and HP Cells, 2nd Gen 7nm, And The Snapdragon 855 DTCO},
    author={David Schor},
    year={2019},
    url={https://fuse.wikichip.org/news/2408/tsmc-7nm-hd-and-hp-cells-2nd-gen-7nm-and-the-snapdragon-855-dtco/}
}

@inproceedings{Gupta2015DeepLW,
  title={Deep Learning with Limited Numerical Precision},
  author={Suyog Gupta and Ankur Agrawal and K. Gopalakrishnan and Pritish Narayanan},
  booktitle={International Conference on Machine Learning},
  year={2015}
}

@article{Venkatesh2016AcceleratingDC,
  title={Accelerating Deep Convolutional Networks using low-precision and sparsity},
  author={Ganesh Venkatesh and Eriko Nurvitadhi and Debbie Marr},
  journal={2017 IEEE International Conference on Acoustics, Speech and Signal Processing (ICASSP)},
  year={2016},
  pages={2861-2865}
}

@inproceedings{Lahiri2013AMF,
  title={A memory frontier for complex synapses},
  author={Subhaneil Lahiri and Surya Ganguli},
  booktitle={NIPS},
  year={2013}
}

@inproceedings{Sandberg2008WholeBE,
  title={Whole Brain Emulation},
  author={Anders Sandberg and Nick Bostrom and James Martin},
  year={2008}
}

@article{Bartol2015NanoconnectomicUB,
  title={Nanoconnectomic upper bound on the variability of synaptic plasticity},
  author={Thomas M. Bartol and Cailey Bromer and Justin P. Kinney and Michael Chirillo and Jennifer N. Bourne and Kristen M. Harris and Terrence J. Sejnowski},
  journal={eLife},
  year={2015},
  volume={4}
}

@article{carlsmith2020brain,
    title={How Much Computational Power Does It Take to Match the Human Brain?},
    author={Joseph Carlsmith},
    year={2020},
    url={https://www.openphilanthropy.org/research/how-much-computational-power-does-it-take-to-match-the-human-brain/}
}

@inproceedings{Swanson1972IonimplantedCM,
  title={Ion-implanted complementary MOS transistors in low-voltage circuits},
  author={Robert M. Swanson and James D. Meindl},
  year={1972}
}

@article{Zhai2005TheLO,
  title={The limit of dynamic voltage scaling and insomniac dynamic voltage scaling},
  author={Bo Zhai and David Blaauw and Dennis Sylvester and Kriszti{\'a}n Flautner},
  journal={IEEE Transactions on Very Large Scale Integration (VLSI) Systems},
  year={2005},
  volume={13},
  pages={1239-1252}
}

@article{Meindl2000TheFL,
  title={The fundamental limit on binary switching energy for terascale integration (TSI)},
  author={James D. Meindl and Jeffrey A. Davis},
  journal={IEEE Journal of Solid-State Circuits},
  year={2000},
  volume={35},
  pages={1515-1516}
}

@article{Zhai2004TheoreticalAP,
  title={Theoretical and practical limits of dynamic voltage scaling},
  author={Bo Zhai and David Blaauw and Dennis Sylvester and Kriszti{\'a}n Flautner},
  journal={Proceedings. 41st Design Automation Conference, 2004.},
  year={2004},
  pages={868-873}
}

@misc{ISO2015nanotechnologies,
    title={ISO/TS 80004-1:2015 Nanotechnologies — Vocabulary — Part 1: Core terms},
    author={International Organization for Standardization},
    year={2015},
    url={https://www.iso.org/standard/68058.html}
}

@article{Jayaprakasan2011EvaluationOT,
  title={Evaluation of the Conventional vs. Ancient Computation Methodology for Energy Efficient Arithmetic Architecture},
  author={V. Jayaprakasan and S. Vijayakumar and V. S. Kanchana Bhaaskaran},
  journal={2011 International Conference on Process Automation, Control and Computing},
  year={2011},
  pages={1-4}
}

@article{Badran2019LowLC,
  title={Low Leakage Current Symmetrical Dual-k 7 nm Trigate Bulk Underlap FinFET for Ultra Low Power Applications},
  author={Mahmoud S. Badran and Hanady Hussein Issa and Saleh M. Eisa and Hani Fikry Ragai},
  journal={IEEE Access},
  year={2019},
  volume={7},
  pages={17256-17262}
}

@inproceedings{Khanna2016ShortChannelEI,
  title={Short-Channel Effects in MOSFETs},
  author={Vinod K. Khanna},
  year={2016}
}

@article{Huang1999Sub5F,
  title={Sub 50-nm FinFET: PMOS},
  author={Xuejue Huang and Wen-Chin Lee and Charles Kuo and Digh Hisamoto and Leland Chang and Jakub Kedzierski and Erik H. Anderson and Hideki Takeuchi and Yang‐Kyu Choi and K. Asano and Vivek Subramanian and Tsu-Jae King and Jeffrey Bokor and Chenming Hu},
  journal={International Electron Devices Meeting 1999. Technical Digest (Cat. No.99CH36318)},
  year={1999},
  pages={67-70}
}

@article{Chen2014DianNaoAS,
  title={DianNao: a small-footprint high-throughput accelerator for ubiquitous machine-learning},
  author={Tianshi Chen and Zidong Du and Ninghui Sun and Jia Wang and Chengyong Wu and Yunji Chen and Olivier Temam},
  journal={Proceedings of the 19th international conference on Architectural support for programming languages and operating systems},
  year={2014}
}

@inproceedings{Zhirnov2014MinimumEO,
  title={Minimum Energy of Computing, Fundamental Considerations},
  author={Victor V. Zhirnov and Ralph K. Cavin and Luca Gammaitoni},
  year={2014},
  url={https://api.semanticscholar.org/CorpusID:30075410}
}

@article{Frank2020ReversibleCW,
  title={Reversible Computing with Fast, Fully Static, Fully Adiabatic CMOS},
  author={Michael P. Frank and Robert W. Brocato and Brian D. Tierney and Nancy A. Missert and Alexander H. Hsia},
  journal={2020 International Conference on Rebooting Computing (ICRC)},
  year={2020},
  pages={1-8},
  url={https://api.semanticscholar.org/CorpusID:221397093}
}

@inproceedings{Abbas2020WiresAC,
  title={Wires and Clocks},
  author={Karim Abbas},
  year={2020},
  url={https://api.semanticscholar.org/CorpusID:213610507}
}

@MISC{shashank2021why,
    TITLE = {Why would an AND gate need six transistors?},
    AUTHOR = {Shashank V M},
    HOWPUBLISHED = {Electrical Engineering Stack Exchange},
    NOTE = {URL:https://electronics.stackexchange.com/q/533539 (version: 2021-06-13)},
    EPRINT = {https://electronics.stackexchange.com/q/533539},
    URL = {https://electronics.stackexchange.com/q/533539}
}

@misc{nvidia2023hopper,
title = {NVIDIA H100 Tensor Core GPU Architecture},
author = {Nvidia},
url = {https://resources.nvidia.com/en-us-tensor-core},
year = {2023}
}

@article{eadline2023nvidia,
  title={Nvidia H100: Are 550,000 GPUs Enough for This Year?},
  author={Eadline, Doug},
  journal={HPCwire},
  url={https://www.hpcwire.com/2023/08/17/nvidia-h100-are-550000-gpus-enough-for-this-year/},
  year={2023},
  month={8},
  day={17}
}

@misc{eia2023power,
title={Average Price of Electricity to Ultimate Customers by End-Use Sector},
url={https://web.archive.org/web/20231117125354/https://www.eia.gov/electricity/monthly/epm_table_grapher.php?t=epmt_5_6_a},
publisher={Energy Information Administration},
author={{Energy Information Administration}},
year={2023},
month={11},
day={17}
}

\end{document}